\documentclass[]{raa}            

\usepackage{graphicx,times}             
\usepackage{natbib}
\usepackage{amssymb,amsmath}
\bibpunct{(}{)}{;}{a}{}{,}

\usepackage[a4paper=true,dvipdfm=true,pagebackref=true]{hyperref}
\hypersetup{colorlinks = true, linkcolor = green, anchorcolor = red, citecolor = blue, filecolor = red, pagecolor = red, urlcolor = red}

\begin{document}

   \title{Photometric analysis and evolutionary stages of the contact binary V2790~Ori
}

   \volnopage{Vol.0 (20xx) No.0, 000--000}      
   \setcounter{page}{1}          

   \author{Wichean Kriwattanawong
      \inst{1,2}
   \and Kriangsak Kriwattanawong
      \inst{3}
   }

   \institute{Department of Physics and Materials Science, Faculty of Science, Chiang Mai University, Chiang Mai, 50200, Thailand; {\it wichean.k@cmu.ac.th}\\
        \and
             Research Center in Physics and Astronomy, Faculty of Science, Chiang Mai University, Chiang Mai, 50200, Thailand\\
        \and
             Department of Chemical Engineering, Faculty of Engineering, King Mongkut's Institute of Technology Ladkrabang, Bangkok 10520, Thailand\\
\vs\no
   {\small Received~~20xx month day; accepted~~20xx~~month day}
}

\abstract{ 
A photometric analysis and evolutionary stages of the contact binary V2790~Ori is presented. The $BVR_\mathrm{C}$ observations were carried out at the Thai National Observatory. The photometric light curves were fitted to provide fundamental parameters, required to use in the analysis of evolutionary stages of the binary. The results show that V2790~Ori is a W-type contact system with a mass ratio of $q$~=~2.932. The orbital period increase is found at a rate of $\mathrm{d}P\slash \mathrm{d}t~=~$1.03$\times10^{-7}~$d~yr$^{-1}$. It implies that a rate of mass transfer from the secondary component to the primary one is $\mathrm{d}m_2\slash \mathrm{d}t~=~$6.31$\times10^{-8}~M_{\odot}$~yr$^{-1}$. Furthermore, we find that from the detached phase to the contact phase, mass of the evolved secondary component has been lost 1.188$\pm$0.110~$\mathrm{M}_{\odot}$, i.e., mass lost by the system of 0.789$\pm$0.073~$\mathrm{M}_{\odot}$ and mass transfer to the primary of 0.399$\pm$0.037~$\mathrm{M}_{\odot}$. Since the time of the first overflow, the angular momentum loss is found to be 72.2~$\%$ of $J_\mathrm{FOF}$, causing the orbit and Roche surface to shrink until the present time. 
\keywords{stars: binaries: close --- stars: binaries: eclipsing --- stars: individual: V2790~Ori}
}

   \authorrunning{Wichean Kriwattanawong \& Kriangsak Kriwattanawong }            
   \titlerunning{Photometric analysis and evolutionary stages of V2790~Ori}  

   \maketitle

%
%
\section{Introduction}           
\label{sect:intro}

Contact binaries are believed to have been formed from detached systems by evolutionary expansion of the components or the angular momentum loss (AML) due to magnetic breaking (e.g. \citealt{VIHU81,RUCI86,JIAN14}). \citet{DEMI06} showed an evidence of decreasing rates of the angular momentum, systematic mass and orbital period of a sample of 114 detached systems, derived from their kinematics. However, the evolution is controlled not only by the AML, but also by mass loss and mass transfer between the components (e.g. \citealt{YAKU05,EKER08}). The evolution is driven by a slow expansion of the progenitor of the secondary component (the evolved component near or after the terminal age main sequence), followed by mass transfer to the other component, accompanied by the AML due to stellar winds and mass loss. The components of close detached binaries approach each other to form contact binaries. The continuous AML and mass loss can bring the components closer together, getting a smaller orbit and shrinking the Roche surface (e.g. \citealt{VIHU81,IBEN84,TUTU04,GAZE08,STEP12}).
 
V2790~Ori [RA(J2000.0) = 06$^{\rm h}$15$^{\rm m}$31$^{\rm s}$.40, Dec.(J2000.0) = +19$^{\circ}$35$^{\prime}$ 22$^{\prime}$$^{\prime}$.1] is a contact binary, found in the Northern Sky Variability Survey (NSVS). Its orbital period was firstly reported by \citet{OTER04} to be 0.287842 d with a primary minimum at HJD~2451521.695. Two years later, \citet{AMMO06} contributed the effective temperature of V2790~Ori to be 5643 K. Furthermore, two recent values of the effective temperature of 5713 K and 5576 K  were cataloged by \citet{MCDO17} and \citet{OELK18}, respectively. The average value was obtained to be about 5644 K.
 
This work presents a photometric analysis and evolutionary stages of the contact binary V2790~Ori. The $BVR_\mathrm{C}$ photometric observations are described in Section~\ref{sect:obs}. Section~\ref{sect:per} presents an analysis of a period change. Section~\ref{sect:lc} explains the results of the light curve fit. Section~\ref{sect:evol} presents evolutionary stages, including mass change throughout its evolution from the detached phase to the contact phase and orbital evolution by the AML from the time of the first overflow to the present time. Finally, the main results are summarized in Section~\ref{sect:conc}.

\section{Observations}
\label{sect:obs}

V2790~Ori was observed by using the 0.5 m reflecting telescope at the Thai National Observatory, Chiang Mai, during three nights in 2015 (January 21--23). An Andor iKon-L-936 CCD camera was equipped on the telescope. Total integration times were 120~s for $B$ band and 60~s for $V$ and $R_\mathrm{C}$ bands. We obtained a total of 699 individual observations for the three filter bands. The $BVR_\mathrm{C}$ differential magnitudes of the binary were measured using TYC 1322-1399-1 [RA(J2000.0) = 06$^{\rm h}$15$^{\rm m}$19$^{\rm s}$.52, Dec.(J2000.0) = +19$^{\circ}$37$^{\prime}$ 07$^{\prime}$$^{\prime}$.6] and TYC~1322-1411-1 [RA(J2000.0) = 06$^{\rm h}$15$^{\rm m}$05$^{\rm s}$.39, Dec.(J2000.0) = +19$^{\circ}$40$^{\prime}$ 43$^{\prime}$$^{\prime}$.3] as comparison and check stars, respectively. The observed magnitudes cover light curves with five minimum light times as shown in Table \ref{tab:o-c}. The tricolor light curves vary about 0.64, 0.59 and 0.56 mag, for the $B$, $V$ and $R_\mathrm{C}$ bands, respectively. Max II is found to be slightly brighter than Max I.

\begin{table}[h!]
\begin{center}
\caption{\label{tab:o-c} Times of minimum light for V2790~Ori. }
\begin{tabular}{@{}lcccc@{}}
\hline
\hline

HJD     &      Min     &   Ref.   &   Epoch   &     (O--C)   \\
\hline

2451521.6950	&	~I	&	~[1]	&		&		\\
2453327.7560	&	II	&	~[2]	&		&		\\
2455520.8205	&	II	&	~[3]	&	~-5291.5	&	~0.0005	\\
2455532.9103	&	II	&	~[2]	&   ~-5249.5	&	~0.0009	\\
2455604.2950	&	II	&	~[4]	&	~-5001.5	&	~0.0009	\\
2455632.3597	&	~I	&	~[4]	&	~-4904.0	&	~0.0010	\\
2455644.3050	&	II	&	~[4]	&	~-4862.5	&	~0.0009	\\
2455896.8827	&	~I	&	~[5]	&	~-3985.0	&	-0.0026	\\
2455902.7850	&	II	&	~[6]	&	~-3964.5	&	-0.0011	\\
2455959.3466	&	~I	&	~[7]	&	~-3768.0	&	-0.0004	\\
2456288.0610	&	~I	&	~[8]	&	~-2626.0	&	-0.0013	\\
2456288.2045	&	II	&	~[8]	&	~-2625.5	&	-0.0018	\\
2456623.1092	&	~I	&	~[9]	&	~-1462.0	&	-0.0010	\\
2456623.2547	&	II	&	~[9]	&	~-1461.5	&	~0.0006	\\
2457041.6326	&	~I	&	[10]	&	~~~~~~~~-8.0	&	~0.0004	\\
2457041.7767	&	II	&	[10]	&	~~~~~~~~-7.5	&	~0.0006	\\
2457042.6403	&	II	&	[10]	&	~~~~~~~~-4.5	&	~0.0007	\\
2457042.7836	&	~I	&	[10]	&	~~~~~~~~-4.0	&	~0.0000	\\
2457044.0792	&	II	&	[11]	&	~~~~~~~~~0.5	&	~0.0004	\\
2457045.0865	&	~I	&	[11]	&	~~~~~~~~~4.0	&	~0.0002	\\
2457045.2306	&	II	&	[11]	&	~~~~~~~~~4.5	&	~0.0004	\\
2457046.0942	&	II	&	[11]	&	~~~~~~~~~7.5	&	~0.0005	\\
2457046.2377	&	~I	&	[11]	&	~~~~~~~~~8.0	&	~0.0000	\\
2457048.6848	&	II	&	[10]	&	~~~~~~~16.5	&	~0.0005	\\
2457048.8286	&	~I	&	[10]	&	~~~~~~~17.0	&	~0.0004	\\
2457049.6919	&	~I	&	[10]	&	~~~~~~~20.0	&	~0.0001	\\
2457049.8362	&	II	&	[10]	&	~~~~~~~20.5	&	~0.0005	\\
2457050.6995	&	II	&	[10]	&	~~~~~~~23.5	&	~0.0003	\\
2457050.8432	&	~I	&	[10]	&	~~~~~~~24.0	&	~0.0001	\\
2457064.3746	&	~I	&	[12]	&	~~~~~~~71.0	&	~0.0029	\\
2457345.7361	&	II	&	[10]	&	~~~~1048.5	&	-0.0010	\\
2457345.8798	&	~I	&	[10]	&	~~~~1049.0	&	-0.0012	\\
2457346.7434	&	~I	&	[10]	&	~~~~1052.0	&	-0.0011	\\
2457346.8876	&	II	&	[10]	&	~~~~1052.5	&	-0.0008	\\
2457384.0202	&	II	&	[13]	&	~~~~1181.5	&	~0.0002	\\
2457384.1636	&	~I	&	[13]	&	~~~~1182.0	&	-0.0003	\\

\hline
\end{tabular}
\end{center}

\it Notes:
\rm Column 1: HJD at light minimum. Column 2: types of minimum. Column 3: references for the sources are as follow: [1] \citet{OTER04}; [2] \citet{DIET11}; [3] \citet{NELS11}; [4] \citet{NAGA12}; [5] \citet{DIET12}; [6] \citet{NELS12}; [7] \citet{HUBS13}; [8] \citet{NAGA13}; [9] \citet{NAGA14}; [10] \citet{MICH16}; [11] This study; [12] \citet{JURY17}; [13] \citet{NAGA16};. Column 4: epoch. Column 5: ($O-C$). 

\end{table}

\section{Period analysis}
\label{sect:per}

The first estimate of the orbital period for the contact binary V2790~Ori was provided by \citet{OTER04}. Until the present day, thirty six eclipse timings, including this work, are available as listed in Table \ref{tab:o-c}. Since the first two times are too far from the others, the last thirty four times were used to calculate the orbital period as shown in Equation (\ref{eq:min}). The revised period is found to be 0.2878418 d.

\begin{equation}
\rm Min.I =  \rm HJD 2457043.9349 (\pm 0.0002)  
             + 0.2878418 (\pm 0.0000001) \times \it E.
\label{eq:min}
\end{equation}

\begin{equation}
(O-C) =  -0.00005 (\pm 0.00021) 
         + 1.67 (\pm 2.45) \times 10^{-7} E 
         + 4.06 (\pm 5.63) \times 10^{-11} E^2.
\label{eq:o-c}
\end{equation}

\begin{figure}[h!]
\centering
\vspace*{.6cm}
\includegraphics[width=0.65\textwidth,height=0.5\textheight,angle=-90]{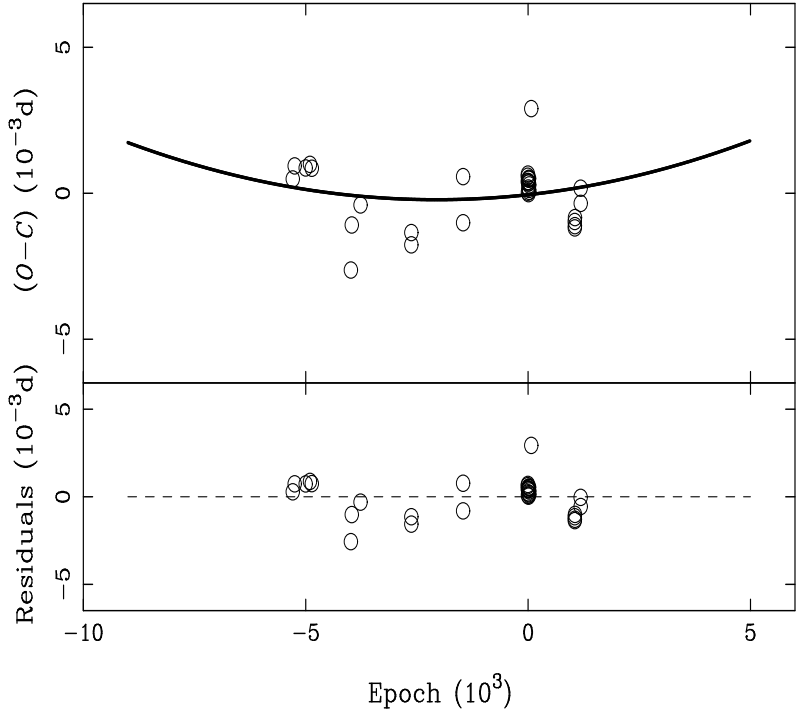}
\caption{($O-C$) curve (upper) and the corresponding residuals (lower) for V2790~Ori.}
\label{fig:o-c}
\end{figure}

With the above ephemeris, a least squares method was used to fit the ($O-C$) residuals as shown in Equation (\ref{eq:o-c}). The ($O-C$) fit yields an upward quadratic curve as shown in Figure \ref{fig:o-c}. It is interpreted for the orbital period increase at a rate of $\mathrm{d}P\slash \mathrm{d}t~=~$1.03$\times10^{-7}~$d~yr$^{-1}$.

\section{Light curve fit}
\label{sect:lc}

\begin{figure}[h!]
\centering
\vspace*{.6cm}
\includegraphics[width=0.6\textwidth,height=0.50\textheight,angle=-90]{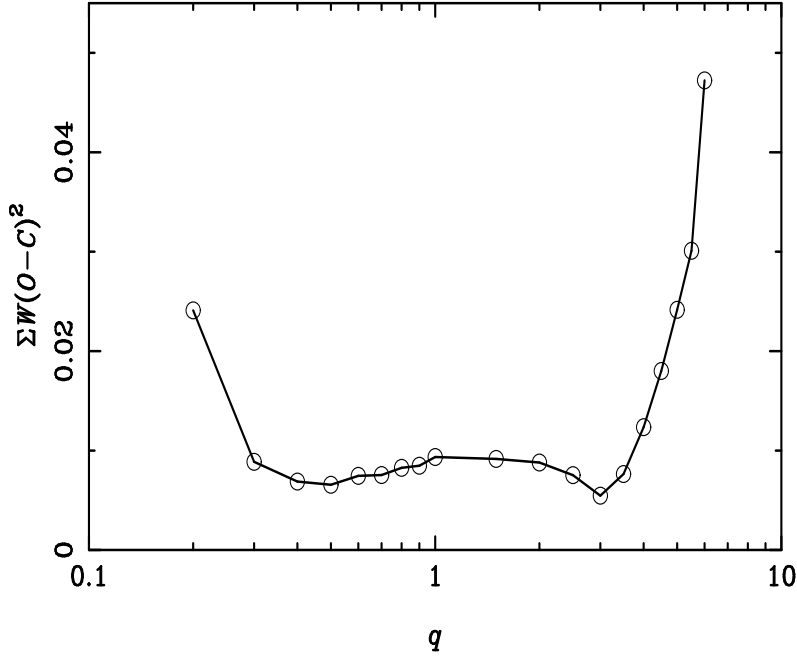}
\caption{$q$-search for V2790~Ori.}
\label{fig:q}
\end{figure}

\begin{figure}[h!]
\centering
\vspace*{.6cm}
\includegraphics[width=0.6\textwidth,height=0.50\textheight,angle=-90]{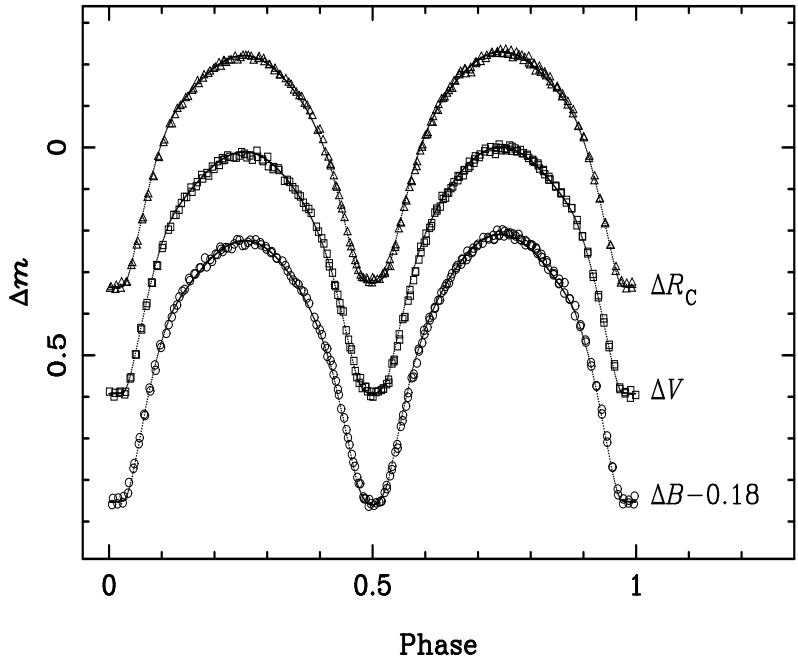}
\caption{Observed light curves in $B$ (circles), $V$ (squares) and $R_\mathrm{C}$ (triangles) filter bands and theoretical light curves (solid lines) versus orbital phase of V2790~Ori.}
\label{fig:lc}
\end{figure}

An analysis of the tricolor light curves for V2790~Ori was done by using the 2013 version of the Wilson-Devinney (W-D) code \citep{WILS71,WILS79,WILS12,WILS10}. According to three available published values of the effective temperature, mentioned in Section \ref{sect:intro}, the average value of 5644 K was assigned to fix the temperature of star 2, $T_2$. The adjustable parameters (the mass ratio, $q$; the effective temperature of star 1, $T_1$; the surface potential of the components, $\Omega_1$ = $\Omega_2$; the orbital inclination, $i$; and the monochromatic luminosities of star 1, $L_1$) were applied to the W-D fit. A spot was added on the star 1. A $q$-search procedure was iterated over a wide range of mass ratio to minimize sum of the squared residuals $\Sigma$$W$($O-C$)$^2$. A good fit is found at about $q\sim$~3 as shown in Figure \ref{fig:q}. With fine iterations, the minimum value of $\Sigma$$W$($O-C$)$^2$ is obtained at $q$~=~2.932$\pm$0.002. The results show that V2790~Ori is a W-type contact binary. Contact configuration of the system is not very deep with a fill-out factor of 20.89$\pm$1.03$\%$. A total eclipse is confirmed by the orbital inclination of 85.1$^{\circ}$. The derived temperature ratio of $T_2\slash T_1$ is 0.964. A cool spot on the star 1 can explain the asymmetric light maxima, well-known as the O'Connell effect. This effect was also similarly found to be evidence of star-spot activities of one component or both components in many W-type contact systems such as V789~Her \citep{LIXI18}, V474~Cam \citep{GUOL18}, RW~Dor \citep{SARO19} and TY~UMa \citep{LIHU15}. The fitted light curves are shown in Figure \ref{fig:lc}. Main parameters are listed in Table \ref{tab:phot}, comparing with a previous work \citep{MICH16}. The mass ratio and the fill-out factor in this study are not very different from the previous work. Other parameters are nearly the same, except the effective temperatures, due to the value of $T_2$ was fixed with different assumptions and source data. However, the temperature ratio is still the same.

\begin{table}[h!]
\begin{center}
\caption{\label{tab:phot} Photometric parameters for V2790~Ori}
\begin{tabular}{@{}lrr@{}}
\hline
\hline

\multicolumn{1}{l}{Parameters } &
\multicolumn{1}{r}{\citet{MICH16}} &
\multicolumn{1}{r}{This study} \\

\hline
$q$ & 3.157($\pm$0.008) & 2.932($\pm$0.002) \\
$T_1$($K$) &   5620($\pm$3) &   5856($\pm$9) \\
$T_2$($K$) &   5471 &   5644\\
$i$ ($^{\circ}$)  & 84.15($\pm$0.20) & 85.1($\pm$0.2) \\
$\Omega_1~=~\Omega_2$ & 6.732($\pm$0.010)  & 6.397($\pm$0.007) \\
$L_{1B}/( L_{1B}~+~L_{2B} )$ & - & 0.3229($\pm$0.0010) \\
$L_{1V}/( L_{1V}~+~L_{2V} )$ & - & 0.3106($\pm$0.0007) \\
$L_{1R_c}/( L_{1R}~+~L_{2R_c} )$ & - & 0.3048($\pm$0.0006) \\
$L_{1g'}/( L_{1g'}~+~L_{2g'} )$ & 0.2966($\pm$0.0007) & - \\
$L_{1r'}/( L_{1r'}~+~L_{2r'} )$ & 0.2867($\pm$0.0005) & - \\
$L_{1i'}/( L_{1i'}~+~L_{2i'} )$ & 0.2830($\pm$0.0005) & - \\
$r_1$ (pole) & - &  0.2798($\pm$0.0009) \\
$r_1$ (side) & - &  0.2929($\pm$0.0011) \\
$r_1$ (back) & - &  0.3330($\pm$0.0021) \\
$r_2$ (pole) & - &  0.4544($\pm$0.0007) \\
$r_2$ (side) & - &  0.4891($\pm$0.0010) \\
$r_2$ (back) & - &  0.5182($\pm$0.0013) \\
$f~(\%$) &     15  &   20.89($\pm$1.03)   \\
$\Sigma$$W$($O-C$)$^2$   & 0.0056  & 0.0035\\
\hline
Spot 1 on star 1 & Hot spot & Cool spot \\
\hline
Spot Colatitude ($^{\circ}$) *  & 105($\pm$5)   & 37($\pm$3) \\
Spot Longitude ($^{\circ}$) *  & 10($\pm$3)  & 264($\pm$4) \\
Spot Radius ($^{\circ}$) *     & 14($\pm$4)    & 26($\pm$2) \\
Temperature Factor *         & 1.16($\pm$0.05)   & 0.90($\pm$0.03) \\
\hline
Spot 2 on Star 2  & Cool spot & \\
\hline
Spot Colatitude ($^{\circ}$) *  & 78($\pm$4)  & -  \\
Spot Longitude ($^{\circ}$) *   & 2($\pm$1)   & -  \\
Spot Radius ($^{\circ}$) *      & 12($\pm$4)  & -  \\
Temperature Factor *            & 0.90($\pm$0.05) & -  \\ 
\hline
 
\end{tabular}
\\


\end{center}
\end{table}

\section{Evolutionary stages}
\label{sect:evol}

\subsection{Mass change}

In order to understand evolutionary status of the contact binary, we firstly calculated masses of both components using correlations between mass, orbital period and mass ratio \citep{GAZE09}. We find that $M_1$~=~0.348$\pm$0.038~$\mathrm{M}_{\odot}$ and $M_2$~=~1.020$\pm$0.112~$\mathrm{M}_{\odot}$ for the secondary and primary components, respectively. According to the orbital period increase with a rate of $\mathrm{d}P\slash \mathrm{d}t~=~$1.03$\times10^{-7}~$d~yr$^{-1}$, found in Section \ref{sect:per}, a mass transfer rate from the less massive hotter component to the other one can be estimated using Equation (\ref{eq:delm}) \citep{SING86,PRIB98}. The mass transfer rate is obtained to be $\mathrm{d}M_2\slash \mathrm{d}t~=~$6.31$\times10^{-8}~\mathrm{M}_{\odot}$~yr$^{-1}$. 

\begin{equation} 
\frac{\dot{P}}{P}  = 3\left(\frac{M_2}{M_1} - 1\right) \frac{\dot{M}_2}{M_2}.  
\label{eq:delm}
\end{equation}

Some W-type contact systems with increasing orbital period were collected to compare the value of $\mathrm{d}P\slash \mathrm{d}t$ as listed in Table \ref{tab:cont}. The period increasing rate for V2790~Ori is found to be a typical value with respect to other W-type systems.  

\begin{table}[h!]
\begin{center}
\caption{\label{tab:cont} W-type contact systems with increasing orbital period.}
\begin{tabular}{@{}lcccccccl@{}}
\hline
\hline

Contact Systems	&	$P$ (d)	&	$q$	&	$i$~($^{\circ}$) 	&	$f~(\%$) 	&	$T_1$($K$)	&	$T_2$($K$) 	&	$\mathrm{d}P\slash \mathrm{d}t~($d~yr$^{-1}$)	&	Ref.* \\
 
\hline
AA UMa	&	0.4681266	&	1.819	&	80.3	&	14.8	&	5965	&	5929	&	4.70~$\times10^{-8}$	&	~~[1]	\\
AB And	&	0.3318911	&	1.786	&	83.2	&	25.2	&	5888	&	5495	&	1.46~$\times10^{-7}$	&	~~[2]	\\
AH Vir	&	0.4075243	&	3.317	&	86.5	&	24.0	&	5671	&	5300	&	2.19~$\times10^{-7}$	&	~~[3]	\\
FI Boo	&	0.3899980	&	2.680	&	38.1	&	50.2	&	5746	&	5420	&	1.65~$\times10^{-7}$	&	~~[4]	\\
TX Cnc	&	0.3828832	&	2.220	&	62.1	&	24.8	&	6537	&	6250	&	3.70~$\times10^{-8}$	&	~~[5]	\\
TY Uma	&	0.3545481	&	2.523	&	84.9	&	13.4	&	6250	&	6229	&	5.18~$\times10^{-7}$	&	~~[6]	\\
UX Eri	&	0.4452823	&	2.681	&	76.9	&	14.0	&	6100	&	6046	&	7.70~$\times10^{-8}$	&	~~[7]	\\
V728 Her	&	0.4712901	&	5.607	&	69.2	&	71.0	&	6787	&	6622	&	3.79~$\times10^{-7}$	&	~~[8]	\\
V1191 Cyg	&	0.3133888	&	9.360	&	80.8	&	57.9	&	6375	&	6215	&	3.13~$\times10^{-6}$	&	~~[9]	\\
V2790 Ori	&	0.2878418	&	2.932	&	85.1	&	21.6	&	5856	&	5644	&	1.03~$\times10^{-7}$	&	[10]			\\
\hline

\end{tabular}
\end{center}

\it Notes:
\rm * References for the sources are as follow: [1] \citet{LEEJ11}; [2] \citet{LIHU14}; [3] \citet{CHEN15}; [4] \citet{CHRI13}; [5] \citet{ZHAN09}; [6] \citet{LIHU15}; [7] \citet{QIAN07}; [8] \citet{YUXI16}; [9] \citet{OSTA14}; [10] This study. 

\end{table}

\begin{figure}[h!]
\centering
\includegraphics[width=0.6\textwidth,height=0.50\textheight,angle=-90]{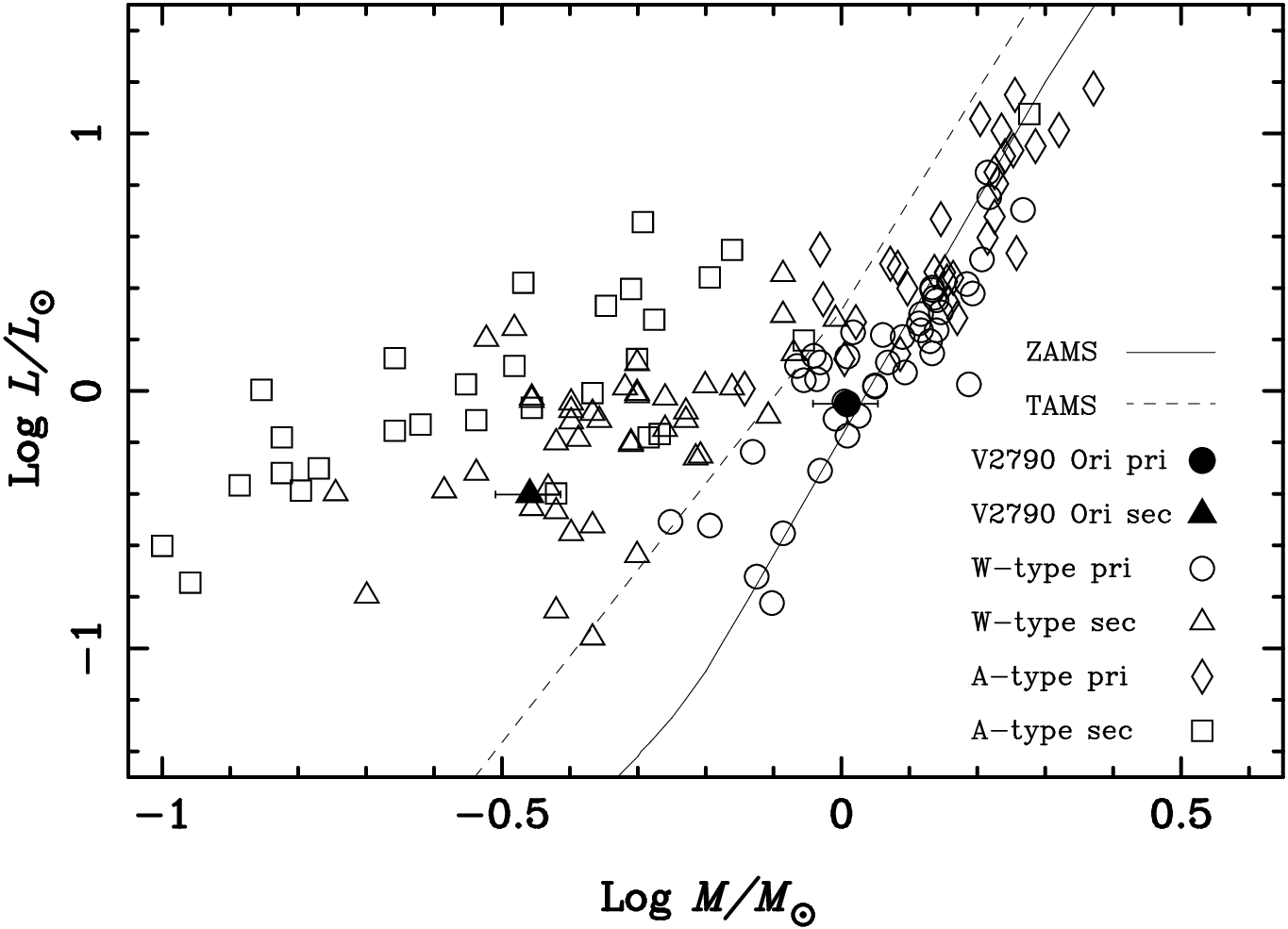}
\includegraphics[width=0.6\textwidth,height=0.50\textheight,angle=-90]{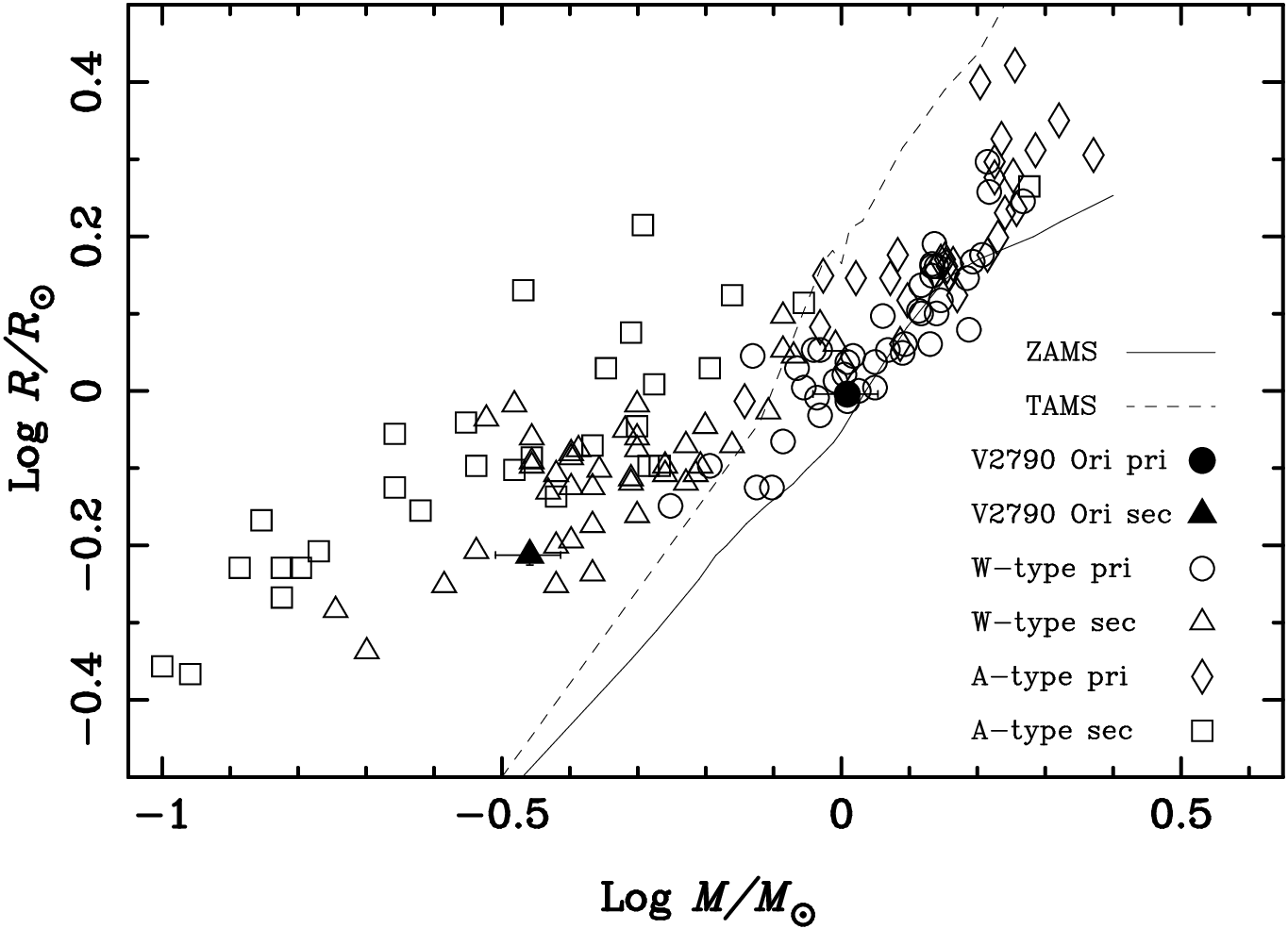}
\caption{Both components of V2790~Ori on the log~$M-$log~$L$ (upper) and log~$M-$log~$R$ (lower) diagrams: the primary more massive (filled dot) and the secondary less massive (filled triangle) components. The sample of contact binaries, collected from the work of \citet{YAKU05} is plotted for comparison. The solid and dashed lines are for the ZAMS and TAMS, respectively, constructed by the \citet{HURL02} binary star evolution code for the solar metallicity. }
\label{fig:evol}
\end{figure}

The photometric parameters in Section \ref{sect:lc} are used to calculate radii and luminosities of the components, and the semimajor axis of the orbit, that are obtained to be $R_1$~=~0.613$\pm$0.018~$\mathrm{R}_{\odot}$, $R_2$~=~0.991$\pm$0.029~$\mathrm{R}_{\odot}$, $L_1$~=~0.396$\pm$0.016~$\mathrm{L}_{\odot}$, $L_2$~=~0.892$\pm$0.036~$\mathrm{L}_{\odot}$ and $a$~=~2.036$\pm$0.059~$\mathrm{R}_{\odot}$, respectively. Masses of both components in the study are slightly greater than values in \citet{MICH16}'s work because of the different method of calculations. In this study, we used the three-dimensional correlations of physical parameters provided by \citet{GAZE09}, while \citet{MICH16} used the mass-period relation supplied by \citet{QIAN03}. Accordingly, our values of $R_1$, $R_2$, $L_1$, $L_2$ and $a$ are slightly greater than the values in the previous work as listed in Table \ref{tab:abs}.

The mass--luminosity and the mass--radius diagrams in Figure \ref{fig:evol} are plotted to compare evolutionary status for both components of V2790~Ori with the zero age main sequence (ZAMS) and the terminal age main sequence (TAMS), constructed by the \citet{HURL02} binary star evolution code. The other well-known contact systems were obtained from the catalog of \citet{YAKU05}. It is found that the primary component of V2790~Ori locates near the ZAMS, similar to the primary stars of other W-type systems, interpreting that its evolutionary stage remains in the main sequence phase. While the secondary component of V2790~Ori lies above the TAMS, indicating that the secondary component has been evolved to be oversized and overluminous. We calculated the mean densities of the components using the equations taken from \citet{MOCH81}:

\begin{equation} 
          \overline{\rho}_1 = \frac{0.079}{V_1(1+q)P^2},
\label{eq:rho1}
\end{equation}

\begin{equation} 
          \overline{\rho}_2 = \frac{0.079 q}{V_2(1+q)P^2} ,
\label{eq:rho2}
\end{equation}

where the relative volumes of the components $V_1$ and $V_2$ are normalized to the semimajor axis, $q$ is the mass ratio, and $P$ is the orbital period. We obtained $\overline{\rho}_1$~=~2.121$\pm$0.259~g~cm$^{-3}$ and $\overline{\rho}_2$~=~1.474$\pm$0.179~g~cm$^{-3}$, that are quite close to the values in \citet{MICH16}'s work. The mean density of the secondary component, $\overline{\rho}_1$, is less than the theoretical value of the ZAMS star, confirming that the component has evolved away from the ZAMS to become oversized. While the mean density of the solar-mass primary star, $\overline{\rho}_2$, is nearly the same value as the Sun, meaning that the primary component is still a main sequence star.

\begin{table}[h!]
\begin{center}
\caption{\label{tab:abs} Absolute parameters for V2790~Ori}
\begin{tabular}{@{}lcr@{}}
\hline
\hline

\multicolumn{1}{@{}l}{Parameters } &
\multicolumn{1}{c}{\citet{MICH16}} &
\multicolumn{1}{r}{This study} \\

\hline
$M_1$~($\mathrm{M}_{\odot}$) & 0.30  & 0.348$\pm$0.038  \\
$M_2$~($\mathrm{M}_{\odot}$) & 0.96  & 1.020$\pm$0.112  \\
$R_1$~($\mathrm{R}_{\odot}$) & 0.58  & 0.613$\pm$0.018  \\
$R_2$~($\mathrm{R}_{\odot}$) & 0.97  & 0.991$\pm$0.029  \\
$L_1$~($\mathrm{L}_{\odot}$) & 0.28  & 0.396$\pm$0.016  \\
$L_2$~($\mathrm{L}_{\odot}$) & 0.68  & 0.892$\pm$0.036  \\
$a$~($\mathrm{R}_{\odot}$)   & 1.98  & 2.036$\pm$0.059  \\ 
$\rho_1$~(g~cm$^{-3}$)   &   2.17   & 2.121$\pm$0.259  \\
$\rho_2$~(g~cm$^{-3}$)   &   1.46   & 1.474$\pm$0.179  \\
log~$g_1$~(cm~s$^{-2}$)   &   4.39   & 4.406$\pm$0.051  \\
log~$g_2$~(cm~s$^{-2}$)   &   4.44   & 4.456$\pm$0.051  \\

\hline
 
\end{tabular}
\\


\end{center}
\end{table}

\begin{table}
\caption{\label{tab:mass} Mass parameters corresponding to the three cases of $\gamma$ for V2790~Ori.}
\begin{center}
\begin{tabular}{@{}lccc@{}}
\hline
\hline

\multicolumn{1}{@{}l}{Mass parameters  } &
\multicolumn{3}{c}{Values~~~~~~} \\ 
\hline
$M_\mathrm{S}$ ($\mathrm{M}_{\odot}$)    &  &   ~0.348($\pm$0.038)  &  \\ 
$M_\mathrm{Si}$ ($\mathrm{M}_{\odot}$)   &  &   ~1.535($\pm$0.148)  &  \\ 
$\Delta M$ ($\mathrm{M}_{\odot}$)        &  &   ~1.188($\pm$0.110)  &  \\ 
$M_\mathrm{P}$ ($\mathrm{M}_{\odot}$)    &  &   ~1.020($\pm$0.112)  &  \\ 
\hline
\multicolumn{1}{l}{ } &
\multicolumn{3}{c}{Values for three cases of $\gamma$} \\
\hline
$\gamma$                        & 0.500 &   ~0.582   & 0.664 \\
$M_\mathrm{Pi}$ ($\mathrm{M}_{\odot}$)   & 0.426($\pm$0.057) &   ~0.523($\pm$0.066)   & 0.620($\pm$0.075) \\ 
$M_\mathrm{Ti}$ ($\mathrm{M}_{\odot}$)   & 1.961($\pm$0.159) &   ~2.058($\pm$0.162)   & 2.156($\pm$0.166) \\ 
$M_\mathrm{lost}$ ($\mathrm{M}_{\odot}$)   & 0.594($\pm$0.055) &  ~0.691($\pm$0.064)   & 0.789($\pm$0.073) \\ 
$M_\mathrm{transfer}$ ($\mathrm{M}_{\odot}$)   & 0.594($\pm$0.055) &   ~0.496($\pm$0.046)   & 0.399($\pm$0.037) \\ 

\hline

\end{tabular}
\end{center}
\end{table}

\begin{figure}[h!]
\begin{center}
\vspace*{.6cm}
\includegraphics[width=0.60\textwidth,height=0.50\textheight,angle=-90]{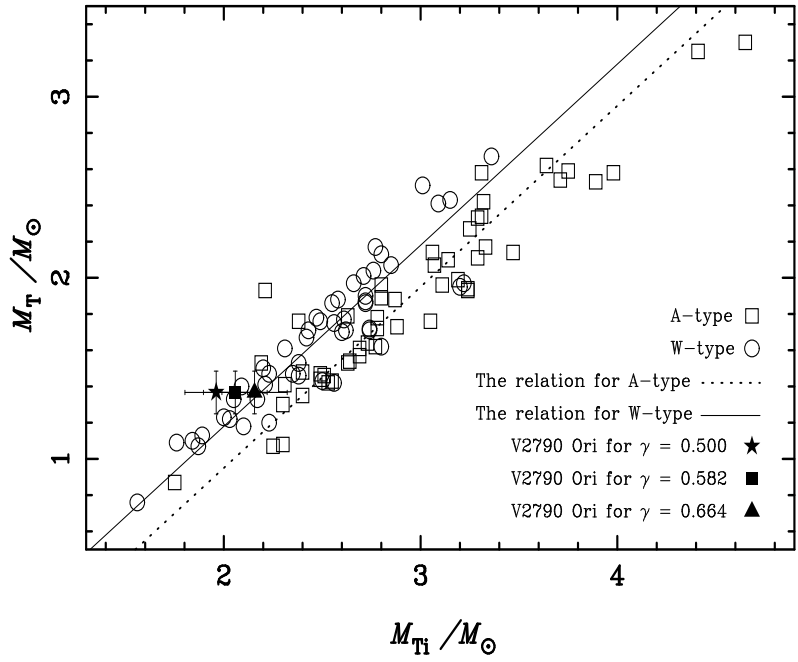}
\vspace*{.6cm}
\end{center}
\caption{Total initial mass of V2790~Ori for the three cases of $\gamma$ on the $M_\mathrm{T}$--$M_\mathrm{Ti}$ diagram. The dotted and solid lines are the $M_\mathrm{T}$--$M_\mathrm{Ti}$ relations, fitted by \citet{YILD14} using a sample of 51 A-type (open squares) and 49 W-type (open circles) contact systems, respectively. V2790~Ori for the value of $\gamma$~=~0.664 (filled triangle) locates closer to the relation than the cases of $\gamma$~=~0.582 (filled square) and $\gamma$~=~0.500 (filled star). }
\label{fig:mtimt}
\end{figure}

With the assumption that mass transfer starts, when the secondary (initially more massive) component has evolved to be near or after the TAMS,  the initial masses of both components were computed from Equations (\ref{eq:MSi})--(\ref{eq:DelM}) \citep{YILD13}:

\begin{equation}
M_\mathrm{Si} = M_\mathrm{S} +\Delta M ,
\label{eq:MSi}
\end{equation}

\begin{equation}
M_\mathrm{Pi} = M_\mathrm{P} - (\Delta M - M_\mathrm{lost}) = M_\mathrm{P} - \Delta M (1 - \gamma),
\label{eq:MPi}
\end{equation}
 
\begin{equation}
\Delta M = 2.50 \left[\left(L_\mathrm{S}/1.49\right)^{1/4.216} - M_\mathrm{S} - 0.07\right]^{0.64},
\label{eq:DelM}
\end{equation}

where $M_\mathrm{Si}$ and $M_\mathrm{Pi}$ are the initial masses, $M_\mathrm{S}$ and $M_\mathrm{P}$ are the current masses of the secondary and the primary stars, respectively. $\Delta M$ is the total mass lost by the secondary, $M_\mathrm{lost}$ is the mass lost by the binary and $\gamma$ is the ratio of $M_\mathrm{lost}$ to $\Delta M$. $L_\mathrm{S}$ is the luminosity of the secondary star. The obtained value of the initial mass of the secondary star is 1.535$\pm$0.148~$\mathrm{M}_{\odot}$ and the mass decrease of the secondary is $\Delta M$~=~1.188$\pm$0.110~$\mathrm{M}_{\odot}$. However, for W-type contact systems, value of the fitting parameter $\gamma$, given by \citet{YILD13} must be in the range of 0.500~$<~\gamma~<~$0.664. Consequently, the corresponding initial mass of the primary star must be a value between 0.426 and 0.620~$\mathrm{M}_{\odot}$, depending on the value of $\gamma$. The precise value of $\gamma$ was needed to assign in the Equation (\ref{eq:MPi}). We applied the minimum, average and maximum values of $\gamma$ of 0.500, 0.582 and 0.664 for three cases to estimate some possible values of the initial mass of the primary star and total initial mass of the binary. The values of mass lost by the system and mass transfer between the components were finally obtained. The possible mass parameters for the three cases are listed in Table \ref{tab:mass}. As shown in Figure \ref{fig:mtimt}, the total initial mass of V2790~Ori for the case of $\gamma$~=~0.664 lies closer to the relation between total present mass ($M_\mathrm{T}$) and total initial mass ($M_\mathrm{Ti}$) for W-type \citep{YILD14} than the other cases. Thus, the appropriate value of $\gamma$ for V2790~Ori should be 0.664. The initial mass of the primary component is obtained to be 0.620$\pm$0.075~$\mathrm{M}_{\odot}$. Since detached phase until the present time, some mass of 0.789$\pm$0.073~$\mathrm{M}_{\odot}$ has been lost from the system. While there could be a mass transfer of 0.399$\pm$0.037~$\mathrm{M}_{\odot}$ from the secondary component to the primary component, after the time of the first overflow (FOF).

\subsection{Orbit and angular momentum change}

\begin{figure}[h!]
\begin{center}
\vspace*{.6cm}
\includegraphics[width=0.60\textwidth,height=0.50\textheight,angle=-90]{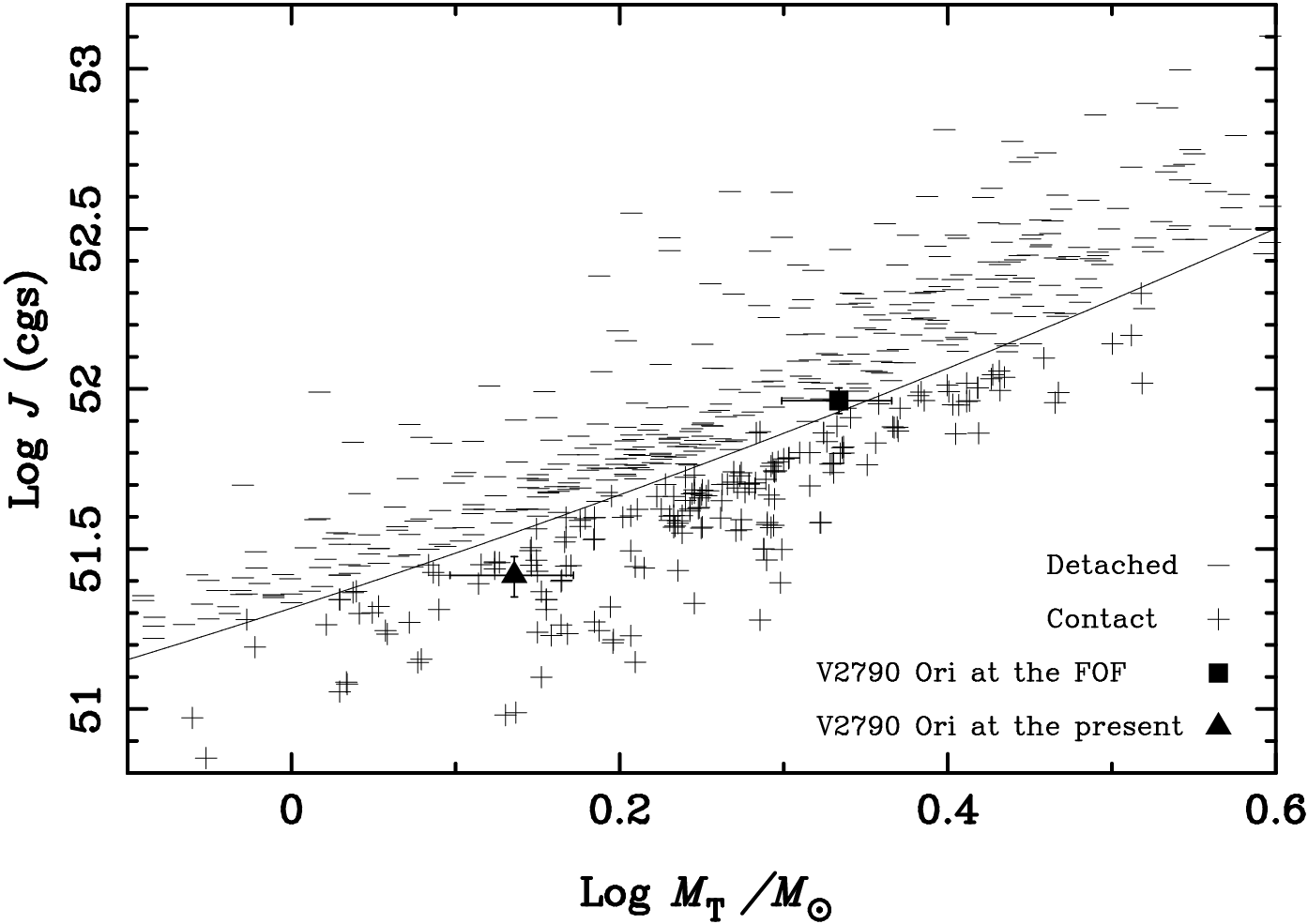}
\vspace*{.6cm}
\end{center}
\caption{Locations of V2790~Ori at the FOF (filled square) and the present time (filled triangle) on the log~$J_\mathrm{o}-$log~$M_\mathrm{T}$ diagram. The samples of detached \citep{EKER06,LEEC15,YILD13} and contact \citep{EKER06,IBAN06} are separated by the quadratic border line \citep{EKER06}.}
\label{fig:jm}
\end{figure}

In general, the orbital angular momentum can be calculated by using the following well known equation:

\begin{equation}
J_\mathrm{o} = \frac{q}{(1+q)^2} \sqrt[3]{\frac{G^2}{2\pi} M_\mathrm{T}^5 P} ,
\label{eq:J}
\end{equation}

where $J_\mathrm{o}$ is the orbital angular momentum, $q$ is the mass ratio, $M_\mathrm{T}$ is the total mass of the binary and $P$ is the orbital period. The value of the present angular momentum of the binary is obtained to be log~$J_\mathrm{o}$~=~51.42$\pm$0.06~cgs. Figure \ref{fig:jm} shows a diagram of the orbital angular momentum versus total mass of contact and detached binaries, separated by the quadratic border line \citep{EKER06}. The sample of detached systems was collected from the catalogs of  \citet{EKER06}, \citet{YILD13} and \citet{LEEC15}. The sample of contact systems was obtained from the works of \citet{EKER06} and \citet{IBAN06}. Location of the binary V2790~Ori for the present time (filled triangle) is found to be below the border line, meaning that the present angular momentum of the binary is less than all detached systems for same mass. It is consistent with that the angular momentum and/or mass loss in the past during the detached phase caused the binary evolved into the contact phase.

During the formation of the contact binary V2790~Ori, the angular momentum loss must continue from detached to semi-detached and contact phases, respectively. The orbital period and semimajor axis decrease lead to the components close together. The angular momentum at the FOF, $J_\mathrm{FOF}$, can be calculated from Equation~(\ref{eq:J}). According to a mass transfer process started at the FOF and the lifetime in the detached phase for W-type is typically negligibly short \citep{YILD14}, we assumed that masses of both components were not much different from the initial values. Thus, for the beginning of the semi-detached phase, initial mass parameters from  Table \ref{tab:mass} were used in the calculation of $J_\mathrm{FOF}$. While the orbital period and the semimajor axis at the FOF was computed using Equations (\ref{eq:PFOF})--(\ref{eq:aFOF}) \citep{YILD14}:

\begin{equation}
P_\mathrm{FOF} = 0.1159 \sqrt{\frac {a_\mathrm{FOF}^{3}}{M_\mathrm{Pi} + M_\mathrm{Si}}} ,
\label{eq:PFOF}
\end{equation}

\begin{equation}
a_\mathrm{FOF} = \frac{0.6 q_\mathrm{i}^{2/3} + \mathrm{ln}(1 + q_\mathrm{i}^{1/3})}{0.49 q_\mathrm{i}^{2/3}} R_\mathrm{TAMS} ,
\label{eq:aFOF}
\end{equation}

where $P_\mathrm{FOF}$ and $a_\mathrm{FOF}$ are the orbital period and the semimajor axis at the FOF, respectively. $R_\mathrm{TAMS}$ is the Roche lobe radius filled by the massive component evolved to reach the TAMS. $P_\mathrm{FOF}$, $a_\mathrm{FOF}$ and $J_\mathrm{FOF}$ are obtained to be 1.083$\pm$0.049 d, 5.731$\pm$0.090~$\mathrm{R}_{\odot}$ and 9.39$(\pm$0.73$)\times10^{51}$ cgs. The results show that the angular momentum has decreased from 9.39$\times10^{51}$ cgs at the FOF to 2.62$\times10^{51}$ cgs at the present time, concurrently with a mass lost by the system of 0.789 $\mathrm{M}_{\odot}$ as shown in Figure \ref{fig:jm}. Consequently, the orbital period and the semimajor axis have reduced from 1.083 d and 5.731~$\mathrm{R}_{\odot}$ to 0.2878420 d and 2.036 $\mathrm{R}_{\odot}$, respectively.

Initially, the binary V2790~Ori in the detached phase consisted of two main sequence stars. The more massive component (the progenitor of the secondary component) has evolved to the TAMS, lead to the resulting oversized envelope. In combination with the AML, the Roche surface was filled by the evolved secondary component, caused the mass transfer to begin. Since the FOF until the present time, the orbit has been reduced by the AML and mass loss.

\section{Conclusions}
\label{sect:conc}

In summary, the $BVR_\mathrm{C}$ observations were carried out at the Thai National Observatory, during three nights in 2015. The photometric data covered five eclipse timings. The ($O-C$) curve shows an orbital period increase with a rate of $\mathrm{d}P\slash\mathrm{d}t~=~$1.03$\times10^{-7}~$d~yr$^{-1}$. The observed light curves were fitted with the W-D method to provide fundamental parameters. It is found that V2790~Ori is a contact system with a mass ratio of $q$~=~2.932. The estimated masses of the components are obtained to be $M_\mathrm{P}$~=~1.020$\pm$0.112~$\mathrm{M}_{\odot}$ and $M_\mathrm{S}$~=~0.348$\pm$0.038~$\mathrm{M}_{\odot}$ for the primary and the secondary, respectively. With an orbital period increase, there could be a mass transfer from the less massive secondary to the more massive primary with a rate of 6.31$\times10^{-8}~\mathrm{M}_{\odot}$~yr$^{-1}$. 

Locations of V2790~Ori's components on the log~$M-$log~$L$ and log~$M-$log~$R$ diagrams confirm that the secondary has evolved to be overluminous and oversized, caused the envelope to fill its Roche surface, while the primary is still a main sequence star. Its mass and angular momentum have been lost continuously throughout its evolution from the detached phase to the contact phase. Since the detached phase until the present time, there could be mass lost by the secondary of 1.188~$\mathrm{M}_{\odot}$, i.e., mass lost by the system of 0.789~$\mathrm{M}_{\odot}$ and mass transfer to the primary of 0.399~$\mathrm{M}_{\odot}$. While the angular momentum has been lost 72.2~$\%$ of $J_\mathrm{FOF}$ from the stage at the FOF to the present time, getting a smaller orbit.

\begin{acknowledgements}
This research work was partially supported by Chiang Mai University. We acknowledge the Thai National Observatory, operated by the National Astronomical Research Institute of Thailand, for the use of the 0.5 m telescope. This work has made use of the SIMBAD online database, operated at CDS, Strasbourg, France and NASA's Astrophysics Data System (ADS), operated by the Smithsonian Astrophysical Observatory (SAO) under a NASA grant.

\end{acknowledgements}

\label{lastpage}

\end{document}